\documentclass[12pt]{article}
\catcode`\@=11
\@addtoreset{equation}{section}

\baselineskip=\normalbaselineskip

\usepackage{mathrsfs,amsbsy,amssymb,latexsym,amsfonts,amsmath}
\usepackage{graphicx,color}

\allowdisplaybreaks

\begin{document}

\begin{titlepage}
\renewcommand{\thefootnote}{\fnsymbol{footnote}}

\begin{flushright}
\parbox{3.5cm}
{KUNS-2265 \\
RIKEN-TH-187}
\end{flushright}

\vspace*{1.5cm}

\begin{center}
{\Large \bf 
Universal description of viscoelasticity \\ 
with foliation preserving diffeomorphisms}%
\footnote{
Based on a talk given at ``Quantum Theory and Symmetries 6 (QTS6)'', 
University of Kentucky, Lexington, KY, USA, July 20--25, 2009}
\end{center}
\vspace{1.5cm}

\centerline{Tatsuo Azeyanagi$^{1}$\footnote{E-mail address: aze@gauge.scphys.kyoto-u.ac.jp},
 Masafumi Fukuma$^{1}$\footnote{E-mail address: 
 fukuma@gauge.scphys.kyoto-u.ac.jp}, 
Hikaru Kawai$^{1,2}$\footnote{E-mail address: 
hkawai@gauge.scphys.kyoto-u.ac.jp}
 and Kentaroh Yoshida$^{1}$\footnote{E-mail address:
 kyoshida@gauge.scphys.kyoto-u.ac.jp}
}

\vspace{2.0cm}

\begin{center}
{\it ${}^{1}$Department of Physics, Kyoto University \\ 
Kyoto 606-8502, Japan\\}
\vspace{0.5cm} 
{\it ${}^{2}$Theoretical Physics Laboratory, Nishina Center, RIKEN\\
Wako, Saitama 351-0198, Japan}
\vspace*{1cm}

\end{center}
\vspace*{1cm}
\begin{abstract}

We review our recent proposal for a universal description 
of generic single-component 
viscoelastic systems with a single relaxation time. 
Foliation preserving diffeomorphisms are introduced as an underlying symmetry 
which naturally interpolates between the two extreme characters 
of elasticity and fluidity. 
The symmetry is found to be powerful enough to determine 
the dynamics in the first order of strains.
\end{abstract}

\begin{flushleft}
\end{flushleft}

\thispagestyle{empty}
\setcounter{page}{0}
\end{titlepage}

\newpage
\renewcommand{\thefootnote}{\arabic{footnote}}
\setcounter{footnote}{0}

\section{Introduction}

Viscoelasticity (see, e.g., \cite{Christensen,Bingham}) 
is a unifying notion which contains solid and fluid as limiting cases. 
It was defined by James Clerk Maxwell  
such as to describe the materials 
that behave as elastic solids at short times scales 
and as viscous fluids at long time scales \cite{LD_fluid, LD_elastic}. 

\begin{figure}[htbp]
\begin{center}
\includegraphics[scale=0.7]{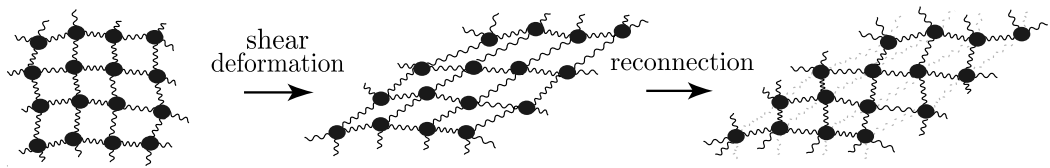}
\caption{Shear deformation of a viscoelastic material 
\label{reconnection}}
\end{center}
\end{figure}

In order to grasp the idea of this definition, 
we consider a material (like a chewing gum) 
consisting of many particles bonding each other. 
They stay at their equilibrium positions in the absence of stresses 
(as in the leftmost of Fig.\ \ref{reconnection}), 
and we now suppose that an external force is applied to deform the material. 
An internal stress is then produced in the body, 
and according to the definition, this internal stress 
can be treated as an elastic force at least during short intervals of time. 
However, if we keep the deformation much longer than some period of time 
(characteristic to each material and to be called the relaxation time), 
then the bonding structure changes to reduce the free energy, 
and the internal stress vanishes eventually. 
The point here is that the two configurations (the center and the rightmost 
of Fig.\ \ref{reconnection}) 
exhibit the same shape (i.e.\ same configurations for the positions of particles) 
but have different bonding structures.

We thus see that the bonding structure itself is dynamical 
for viscoelastic materials, 
and is lead to introduce a new dynamical variables 
(to be called {\it intrinsic metric} \cite{afky}) 
to specify the bonding structure. 
It was claimed in \cite{afky} 
that viscoelasticity may have a universal description 
if we define the intrinsic metric properly 
and assume the invariance of the system 
under {\it foliation preserving diffeomorphisms} (FPDs). 
In fact, it was shown there that FPDs naturally interpolate between the two extreme limits 
of elasticity and fluidity, 
and are powerful enough to determine the dynamics in the first order of strains. 

The main purpose of the present article is 
to give a brief review of \cite{afky} in a concise way, 
and also to present some additional stuffs which were not mentioned in \cite{afky}. 

\begin{figure}[htbp]
\begin{center}
\includegraphics[scale=0.7]{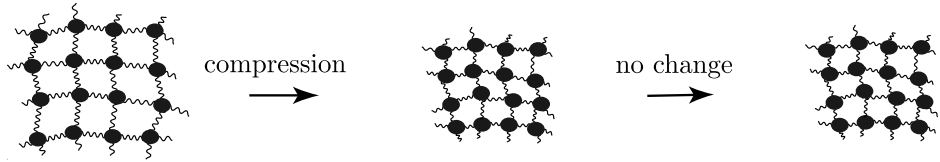}
\caption{Compression of a viscoelastic material
\label{bulk}}
\end{center}
\end{figure}

Our discussions will be constrained to single-component materials. 
Note that, even for this case, there can be two relaxation times, 
one being for shear deformations 
and another for bulk compressions (or expansions). 
In this article, we set the relaxation time for bulk compressions to be infinite, 
so that the bulk stress does not undergo relaxations (see 
Fig.\,\ref{bulk}). 
This ensures that the system exhibits fluidity at long intervals of time.

\section{Geometrical setup}
\label{geometricsetup}

We describe a viscoelastic material 
within the framework of world-volume theory 
with the world-volume coordinates
$ \xi=(\xi^a)$ $(a=1,2,3) $
(labeling the material particles of a body) 
and with the target space coordinates
$X=(X^i)$ $(i=1,2,3)$.
The shape of the material at time $t$ is then specified 
by giving the functions $X^i(\xi,t)$.
In order to further describe the bonding structure, 
we introduce an intrinsic metric $\bar{h}_{ab}(\xi,t)$ as follows.

\begin{figure}[htbp]
\begin{center}
\includegraphics[scale=0.5]{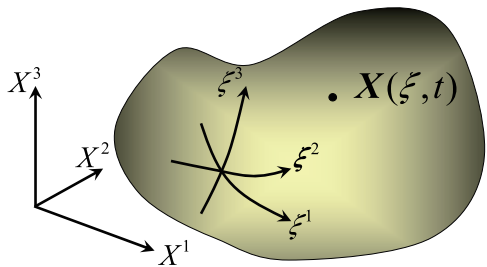}
\caption{World-volume of a viscoelastic material
\label{world-volume}}
\end{center}
\end{figure}  

\begin{figure}[htbp]
\begin{center}
\includegraphics[scale=0.5]{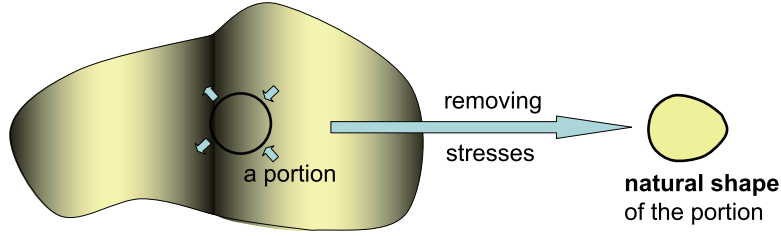}
\caption{Definition of the natural shape for a small portion
\label{natural_shape}}
\end{center}
\end{figure}

We first recall that our viscoelastic materials are defined 
to be elastic at least for a very short time. 
This means that for any (sufficiently small) portion in a material, 
its {\it natural shape} can be defined at each moment 
as the shape that would be taken when all stresses were virtually removed 
(see Fig.\ \ref{natural_shape}).

\begin{figure}[htbp]
\begin{center}
\includegraphics[scale=0.5]{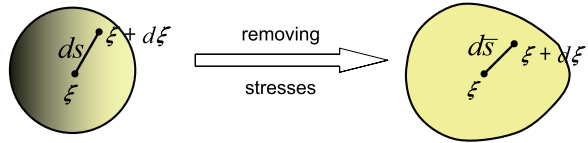}
\caption{The induced and intrinsic metrics
\label{intrinsic_metric}}
\end{center}
\end{figure}

This observation enables us to define two kinds of metrics 
on the world-volume at each time $t$ (see Fig.\ \ref{intrinsic_metric}). 
The first one is the {\it induced metric} $h_{ab}(\xi,t)$ 
which measures the length in the real shape: 
\begin{align}
 ds^2=h_{ab}(\xi,t)\,d\xi^a d\xi^b\,,
\end{align}
where 
\begin{align}
 h_{ab}(\xi,t)\equiv e^i_a(\xi,t)e^i_b(\xi,t)
  \qquad \bigl( e^i_a(\xi,t)\equiv \partial_a X^i(\xi,t)\,:~\mbox{dreibein}\bigr)\,.
\end{align}
Another is the {\it intrinsic metric} $\bar{h}_{ab}(\xi,t)$  
which measures the length in the natural shape: 
\begin{align}
 d\bar{s}^2=\bar{h}_{ab}(\xi,t)\,d\xi^a d\xi^b\,. 
\end{align}
The difference between them defines the {\it strain tensor}, 
\begin{align}
 \varepsilon_{ab}(\xi,t)\equiv \frac{1}{2}\bigl( h_{ab}(\xi,t)-\bar{h}_{ab}(\xi,t)\bigr).
\end{align}
In fact, for an elastic material at rest without stress, 
we can take the coordinates $\xi=(\xi^a)$ such that $\bar{h}_{ab}=\delta_{ab}$. 
Then, when the material is slightly deformed under stresses, 
$X^a(\xi,t)=\xi^a+u^a(\xi,t)$ ($u^a$: displacements), 
we have 
$h_{ab}=\partial_a X^i \partial_b X^i=\delta_{ab}+\partial_a u_b +\partial_b u_a +O(u^2)$, 
so that $\varepsilon_{ab}=(1/2)\bigl(\partial_a u_b + \partial_b u_a\bigr)+O(u^2)$.

We denote by $\rho_0$ the mass density in the absence of strains.
In this article, we assume that $\rho_0$ is a constant independent 
of $\xi$ and $t$. 
Then the mass contained in a volume element $d^3\xi=d\xi^1 d\xi^2 d\xi^3$ 
is given by $\displaystyle \rho_0\sqrt{\bar{h}} \,d^3\xi$. 
The mass density $\rho$ in the real three-dimensional space is given 
by the mass conservation $\displaystyle \rho\sqrt{h}d^3\xi = \rho_0 \sqrt{\bar{h}}d^3\xi$ 
as 
$
 \rho = \bigl(\displaystyle\sqrt{\bar{h}}/\sqrt{h}\bigr)\,\rho_0
$.

Let us consider various deformations of a 2D material. 
Here $(\xi^1,\xi^2)$ are comoving coordinates attached to material particles. 

\begin{figure}[htbp]
\begin{center}
\includegraphics[scale=0.55]{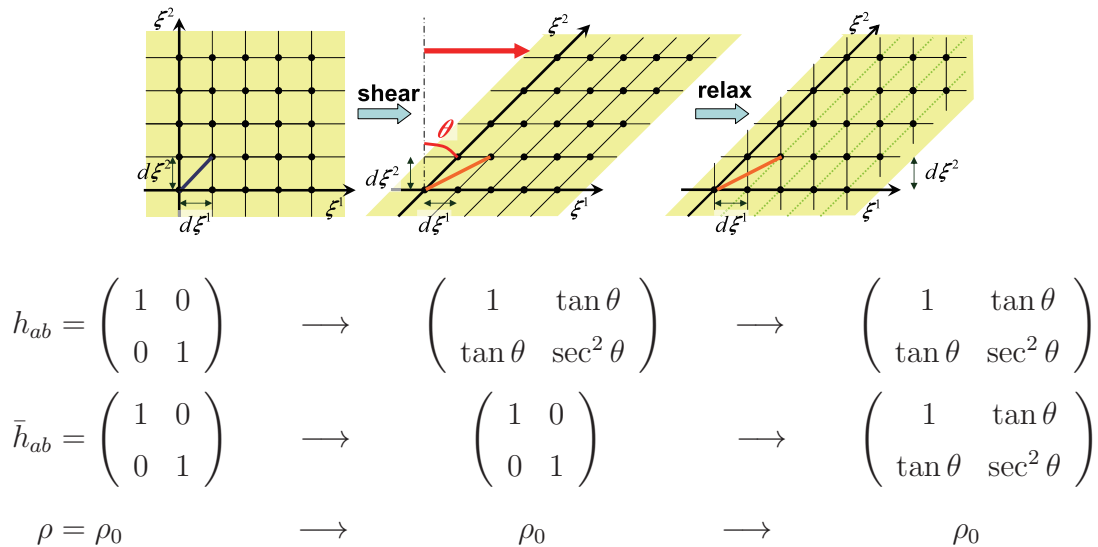}
\begin{align}
h_{ab} &= \left(
  \begin{array}{cc}
    1   & 0   \\
    0  & 1  \\
  \end{array}
\right)  
\qquad
\longrightarrow 
\qquad
\left(
  \begin{array}{cc}
    1   &  \tan\theta   \\
    \tan\theta   & \sec^2\theta   \\
  \end{array}
\right) 
\qquad
\longrightarrow 
\qquad
\left(
  \begin{array}{cc}
    1   &  \tan\theta   \\
    \tan\theta   & \sec^2\theta   \\
  \end{array}
\right) 
\nonumber\\
\bar{h}_{ab} &= \left(
  \begin{array}{cc}
     1  &  0  \\
     0  &  1  \\
  \end{array}
\right) 
\qquad
\longrightarrow
\qquad\,\quad\,\,
 \left(
  \begin{array}{cc}
     1  &  0  \\
     0  &  1  \\
  \end{array}
\right) 
\qquad\qquad
\longrightarrow
\qquad
 \left(
  \begin{array}{cc}
     1  &  \tan\theta  \\
     \tan\theta  &  \sec^2\theta  \\
  \end{array}
\right) 
\nonumber \\
\rho&=\rho_0 \nonumber
\quad\qquad\quad~~~~
\longrightarrow
\qquad\qquad\quad\,
\rho_0 
\qquad\qquad~~~~
\longrightarrow 
\qquad\qquad\quad\,\,\,
\rho_0 \nonumber
\end{align}
\caption{Shear deformation with the angle $\theta\,$
\label{shear_deform}}
\end{center}
\end{figure}

\noindent
\underline{\bf (a) shear deformations:} 
Suppose that a material in equilibrium is deformed with an external force 
as in Fig.\ \ref{shear_deform}. 
The two metrics $h_{ab}$ and $\bar{h}_{ab}$ agree before the deformation. 
When the material is deformed, the induced metric $h_{ab}$ varies, 
expressing the change of the shape. 
In contrast, the intrinsic metric $\bar{h}_{ab}$ retains its original form 
just after the deformation. 
This is because the material should exhibit elasticity for a short period of time, 
and the natural shape at that moment is its original form in equilibrium.  
At sufficiently later times, $\bar{h}_{ab}$ gradually changes its values 
and comes to equal the induced metric.

\begin{figure}[htbp]
\begin{center}
\includegraphics[scale=0.55]{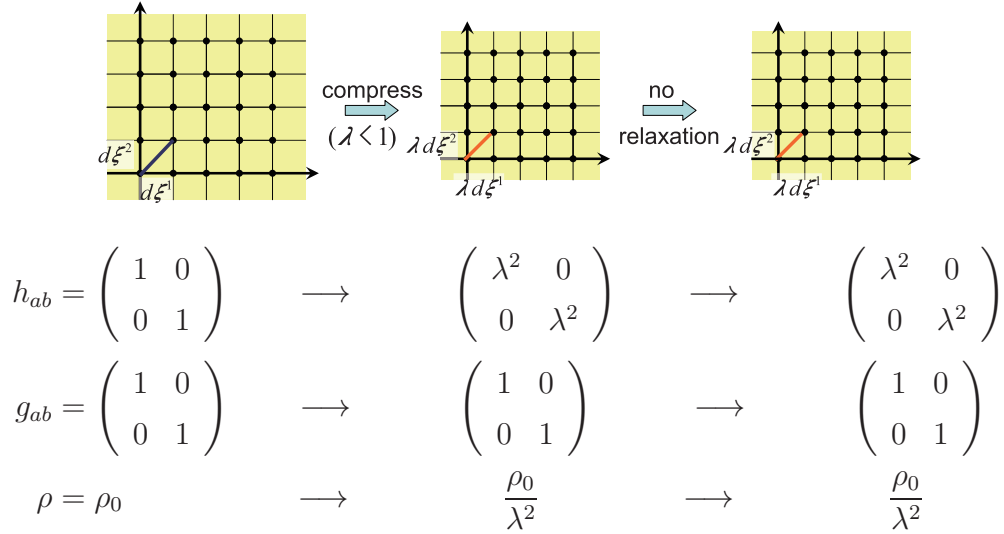}
\begin{align}
h_{ab} &= \left(
  \begin{array}{cc}
    1   & 0   \\
    0  & 1  \\
  \end{array}
\right)  
\qquad
\longrightarrow 
\qquad\quad
\left(
  \begin{array}{cc}
    \lambda^2   &  0   \\
    0  & \lambda^2   \\
  \end{array}
\right) 
\qquad
\longrightarrow 
\qquad\quad
\left(
  \begin{array}{cc}
    \lambda^2 &  0   \\
    0  & \lambda^2   \\
  \end{array}
\right) 
\nonumber\\
g_{ab} &= \left(
  \begin{array}{cc}
     1  &  0  \\
     0  &  1  \\
  \end{array}
\right) 
\qquad
\longrightarrow
\qquad\,\quad
 \left(
  \begin{array}{cc}
     1  &  0  \\
     0  &  1  \\
  \end{array}
\right) 
\qquad\quad
\longrightarrow
\qquad\quad\,
 \left(
  \begin{array}{cc}
     1  &  0  \\
     0  &  1  \\
  \end{array}
\right) 
\nonumber \\
\rho&=\rho_0 \nonumber
\qquad\quad~~~~~~~
\longrightarrow
\qquad\qquad~~
\frac{\rho_0}{\lambda^2} 
\qquad\quad~~~~
\longrightarrow 
\qquad\qquad~~
\frac{\rho_0}{\lambda^2} \nonumber
\end{align}
\caption{Compression by the factor $\lambda$ $(\lambda <1)\,$
\label{bulk_deform}}
\end{center}
\end{figure}

\noindent
\underline{\bf (b) bulk compressions:} 
Suppose that a material in equilibrium is compressed with a factor $\lambda\,(<1)$  
as in Fig.\ \ref{bulk_deform}. 
The two metrics $h_{ab}$ and $\bar{h}_{ab}$ again agree before the deformation, 
and the induced metric $h_{ab}$ changes after the deformation. 
However, according to our assumption that the bulk compression (or expansion) 
does not undergo relaxation, 
the intrinsic metric $\bar{h}_{ab}$ retains its original form.

\section{Foliation preserving diffeomorphisms}

\begin{figure}[htbp]
\begin{center}
\includegraphics[scale=0.47]{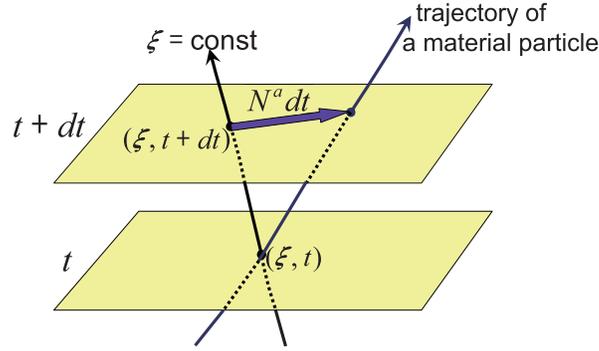}
\caption{Definition of $N^a$ \label{shift_vector}}
\end{center}
\end{figure}

So far the world-volume coordinates $\xi$ have been 
attached to a material, labeling the material particles. 
However, we can introduce arbitrary coordinates $\xi$ at each time $t$, 
independently of the labeling of material particles. 
In this case we need to specify the relative motion of the material 
to the coordinates,  
but this can be carried out by introducing a vector field 
${\cal N}=N^a(\xi,t)\partial_a$ 
(see Fig.\ \ref{shift_vector}) \cite{afky}. 
More precisely, the vector field is defined 
such that the material particle located at $\xi$ at time $t$ 
is supposed to move to the position 
$\xi^a+N^a(\xi,t)d t$ after the time interval $d t$.

With $N^a(\xi,t)$, the system comes to have the gauge symmetry 
of {\it foliation preserving diffeomorphisms} (FPDs), 
reflecting the redundancy in the description. 
The set of FPDs consist of the $(3+1)$-dimensional reparametrizations 
that do not move the time slices, 
and is generated by the infinitesimal transformations of the form 
(see \cite{afky} for references of mathematical details):
\begin{align}
 \delta\xi^a(\xi,t)=\epsilon^a(\xi,t)\,,\quad \delta t(\xi,t)=0\,.
\end{align}
The fields $X^i$, $h_{ab}$ and $\bar{h}_{ab}$ are intrinsically 
three-dimensional and thus transform covariantly as
\begin{align}
 \delta X^i &= \epsilon^a\,\partial_a X^i\,,\\
 \delta h_{ab} &= \partial_a\epsilon^c\,h_{cb}+\partial_b\epsilon^c\,h_{ac}
  +\epsilon^c\,\partial_c h_{ab} \,,\\
 \delta \bar{h}_{ab} &= \partial_a\epsilon^c\,\bar{h}_{cb}
  +\partial_b\epsilon^c\,\bar{h}_{ac}
  +\epsilon^c\,\partial_c \bar{h}_{ab} \,,
\end{align}
while the transformation of the shift vector has an additional inhomogeneous term as
\begin{align}
 \delta N^a = -\partial_c\epsilon^a\,N^c + \epsilon^c\,\partial_c N^a - \dot{\epsilon}^a\,.
\end{align}
The time derivatives of a FPD-covariant tensor $T^{b_1 b_2\cdots}_{a_1 a_2\cdots}$ 
are no longer covariant, 
but can be made so by introducing the covariant derivative \cite{afky}: 
\begin{align}
 \frac{D}{D t}T^{b_1 b_2\cdots}_{a_1 a_2\cdots}
  \equiv \frac{\partial}{\partial t}T^{b_1 b_2\cdots}_{a_1 a_2\cdots}
         + {\cal L}_{\cal N}T^{b_1 b_2\cdots}_{a_1 a_2\cdots}\,,  
\end{align}
where ${\cal L}_{\cal N}$ is the Lie derivative 
with respect to ${\cal N}=N^a\partial_a$.

For example, the velocity $v^i(\xi,t)$ and the acceleration 
$a^i(\xi,t)$ of the material particle located at $(\xi,t)$ are given, respectively, by
\begin{align}
 v^i= \frac{DX^i}{Dt} = \dot{X}^i+N^a\partial_a X^i\,,\qquad
 a^i= \frac{Dv^i}{Dt}= \frac{D^2 X^i}{Dt^2} = \dot{v}^i+N^a\partial_a v^i\,.
\end{align}
This can be easily confirmed by noting that the material particle at $(\xi,t)$
moves after the time interval $d t$ to the point $(\xi+N d t,\,t+ d t)$:
\begin{align}
 X^i(\xi+Ndt,\,t+dt) &= X^i(\xi,t)+
  \bigl[ \dot{X}^i(\xi,t)+N^a(\xi,t)\,\partial_a X^i(\xi,t)\bigr]dt \equiv X^i(\xi,t)+v^i(\xi,t)\,dt\,.
\end{align}

As for the metrics $h_{ab}$ and $\bar{h}_{ab}$, 
their covariant derivatives are given, respectively, by 
\begin{align}
 K_{ab} &\equiv \frac{1}{2}\,\frac{Dh_{ab}}{Dt}=
  \frac{1}{2}\,\bigl(
  \dot{h}_{ab}+\partial_a N^c\,h_{cb}
  +\partial_b N^c\,h_{ac}
  +N^c\,\partial_c h_{ab}\bigr)\label{Kg_ab}\,,
\\
 \bar{K}_{ab} &\equiv \frac{1}{2}\,\frac{D\bar{h}_{ab}}{Dt} 
  =\frac{1}{2}\,\bigl(
  \dot{\bar{h}}_{ab}+\partial_a N^c\,\bar{h}_{cb}
  +\partial_b N^c\,\bar{h}_{ac}
  +N^c\,\partial_c \bar{h}_{ab}\bigr)\label{Kh_ab}\,,
\end{align}
and are called the extrinsic curvatures. 
Note that 
\begin{align}
  K_{ab} - \bar{K}_{ab} = \frac{D \varepsilon_{ab}}{Dt}\,,
\label{khabminuskab}
\end{align}
and
\begin{align}
 K_{ab}=\frac{1}{2}\bigl( \nabla_a v_b + \nabla_b v_a \bigr)\,,
\label{Kh_abgeneral}
\end{align}
where $v_a\equiv e^i_a v^i$ and $\nabla_a$ is the covariant derivative 
with respect to $h_{ab}$ \cite{afky}.

As was stated above, 
the set of FPDs forms a gauge symmetry group of the system, 
and expresses a redundancy in the description of the system. 
This can be gauge-fixed arbitrarily 
according to convenience in describing the dynamics of a given system. 
Two of the useful gauge fixings are the followings:

\medskip

\noindent
{\bf [\,Comoving frame\,]} 
This is useful in describing elastic materials and 
is defined by the condition  $N^a(\xi,t)=0$\,,
which leads to $v^i(\xi,t)=\dot{X}^i(\xi,t)$.

\medskip

\noindent
{\bf [\,Laboratory frame\,]} 
This is useful in describing fluids 
and is defined by the condition $X^a(\xi,t)\equiv \xi^a$,
from which follow the equations 
$v^a=N^a$, $h_{ab}=\delta_{ab}$, $\rho=\rho_0\displaystyle\sqrt{\bar{h}}$,
and $a^a=\dot{v}^a+v^b\,\partial_b v^a$. 
The last equation shows that the covariant derivative in this frame 
becomes the material (Lagrangian) derivative.

\section{Fundamental equations}

We now write down a set of equations 
which determine the time evolution of twelve variables 
$X^i$ (three), $\bar{h}_{ab}$ (six) 
and $N^a$ (three) up to FPDs (three). 
Our strategy is to first write down equations which hold in both of the elastic 
and fluid limits (this can be done in specific frames), 
and then to make the equations covariant under FPDs. 

\subsection{Equation of motion for $\bar{h}_{ab}(\xi,t)$}

We first consider the limiting case of elasticity in the comoving frame ($N^a=0$). 
We then have $\dot{\bar{h}}_{ab}=0$ because the natural shape 
does not evolve in time for elastic materials. 
Its FPD-covariant expression is $\bar{K}_{ab}=0$, 
and thus we expect that nonvanishing $\bar{K}_{ab}$ represents 
genuinely plastic (i.e.\ nonelastic) deformations. 
However, one can show that its trace part $\bar{K}\equiv \bar{h}^{ab}\,\bar{K}_{ab}$ 
always vanishes due to the mass conservation \cite{afky}:
\begin{align}
 \bar{K}=0\,.
\label{mass_conserv}
\end{align}
In fact, in the laboratory frame, $\bar{K}$ is expressed 
as $(1/\rho)\bigl(\dot{\rho}+\partial_a(\rho v^a)\bigr)$  \cite{afky}
and thus must vanish. 
 
The above consideration shows that only the traceless (shear) component 
of the extrinsic curvature 
$\bar{K}_{ab}$ expresses the changing rate of the bonding structure, 
which should be proportional to shear strains in the following form \cite{afky}:   
\begin{align}
 \bar{K}_{ab}-\frac{1}{3}\bar{K}\,\bar{h}_{ab} 
  = \frac{1}{\tau}\Bigl( \varepsilon_{ab} 
     - \frac{1}{3}\bigl(\bar{h}^{cd}\varepsilon_{cd}\bigr)\bar{h}_{ab}\Bigr)\,.
\label{rheology_eq}
\end{align}
Note that the second term on the left-hand side can actually be omitted because $\bar{K}=0$. 
The inverse of the proportional coefficient, $\tau$, has the dimension of time, 
and should correspond to the relaxation time.

In order to check the consistency of this equation (named the {\it rheology equation} 
in \cite{afky}), we take two extreme limits. 
In the elastic limit ($\tau\to\infty$), we obtain $\bar{K}_{ab}=0$, 
which again means that there occurs no time evolution of intrinsic metric. 
In contrast, in the fluid limit ($\tau\to 0$), 
we have $\varepsilon_{ab}\propto\bar{h}_{ab}$, or $h_{ab}\propto\bar{h}_{ab}$. 
This means that the material acquires only homogeneous compression around each point, 
which is actually the characteristic property defining fluid. 

\subsection{Equation of motion for $X^i(\xi,t)$}
The dynamics of $X^i(\xi,t)$ should be expressed as Euler's equation 
\begin{align}
 \rho\,a_a= -\nabla^{b} T_{ba}\,.
\label{Euler_eq}
\end{align}
The leading form of the stress tensor $T_{ab}$ in the derivative 
expansion can be determined 
by the following requirements:
\begin{itemize}
\item $T_{ab}$ be symmetric and covariant under FPDs.
\item $T_{ab}$ be linear in the strain $\varepsilon_{ab}$ and 
the spatial derivative of the velocity, $\nabla_{a} v_b$.
\end{itemize}
These requirements 
imply that $T_{ab}$ is a linear combination of the irreducible 
components of $\varepsilon_{ab}$ 
and $K_{ab}=(1/2)(\nabla_{a}v_b+\nabla_{b}v_a)$ 
(see \eqref{Kh_abgeneral}): 
\begin{align}
 T_{ab} = -2\mu\, {\varepsilon}^{\,\prime}_{ab}
  -\frac{1}{\kappa}\,\varepsilon\,h_{ab}-2\gamma\,{K}'_{ab}
  -\zeta \,K h_{ab}\,,
\label{stress_general}
\end{align}
where
\begin{align}
 \varepsilon \equiv h^{ab}\varepsilon_{ab}\,,\quad 
  {\varepsilon}^{\,\prime}_{ab} \equiv \varepsilon_{ab}
  -\frac{1}{3}\,\varepsilon\, h_{ab}\,, \quad
 K \equiv h^{ab}K_{ab}\,,\quad~~
  {K}'_{ab} \equiv K_{ab}-\frac{1}{3}\,Kh_{ab}\,.
\end{align}

The set of equations (\ref{mass_conserv}), (\ref{rheology_eq}), (\ref{Euler_eq}) 
and (\ref{stress_general}) give nine independent equations, 
and thus can describe the time evolution of twelve dynamical variables 
up to FPDs. 
In order to see the meaning of the coefficients in (\ref{stress_general}), 
we consider two extreme limits of 
elasticity and fluidity. 
In the elastic limit, we have $\bar{K}_{ab}=0$ and thus 
${K}'_{ab}=\bigl(D\varepsilon_{ab}/Dt\bigr)\bigl(1+O(\varepsilon)\bigr)$, 
$K=\bigl(D\varepsilon/Dt\bigr)\bigl(1+O(\varepsilon)\bigr)$. 
Then the stress tensor becomes
\begin{align}
 T_{ab} =  -2\mu\, {\varepsilon}^{\,\prime}_{ab}
  -\frac{1}{\kappa}\,\varepsilon\,h_{ab}-2\gamma\,\frac{D {\varepsilon}^{\,\prime}_{ab}}{Dt}
  -\zeta \,\frac{D\varepsilon}{Dt} h_{ab}
  + O(\varepsilon^2)\,,
\end{align}
which shows that the four terms represent, respectively, 
shear and bulk moduli, and shear and bulk frictions. 
On the other hand, 
the fluid limit is realized by considering at time scales 
much longer than the relaxation time $\tau$. 
By using the rheology equation in the form 
${\varepsilon}^{\,\prime}_{ab}
=\tau\,{K}'_{ab}\bigl(1+O(\varepsilon)\bigr)$, 
Eq.\ (\ref{stress_general}) becomes
\begin{align}
 T_{ab}= -2\eta\, {K}'_{ab}-\zeta \,K h_{ab}
  -\frac{1}{\kappa}\,\varepsilon\,h_{ab} + O(\varepsilon^2)\,,
\end{align}
where $\eta=\gamma+\mu\tau$ and $\zeta$ are shear and bulk viscosities. 
The last term expresses the pressure with $\kappa$ being the bulk compressibility.

\section{Dispersion relations}

In order to get dispersion relations for the set of equations 
(\ref{mass_conserv})--(\ref{stress_general}), 
we expand solutions around the hydrostatic solution 
in the laboratory frame $(X^a(\xi,t)=\xi^a,~h_{ab}=\delta_{ab})$ as
\begin{align}
 \bar{h}_{ab}=\delta_{ab}+\Delta\bar{h}_{ab}\,,\quad
  N^a=0+\Delta N^a\,.
\end{align}
We consider plane waves propagating in the $X^3(=\xi^3)$-direction:   
\begin{align}
 \Delta \bar{h}_{ab}(\xi,t)=\Delta\bar{h}_{ab}(k,\omega)e^{ik\xi^3-i\omega t}\,,\quad
  \Delta N^a(\xi,t)=\Delta N^a(k,\omega)e^{ik\xi^3-i\omega t}\,.
\end{align}
The equations can be separated according to the helicity, 
and we have the following equations:

\medskip

\noindent
\textbullet~\underline{helicity 0 (shear modes)} 
$\bigl\{\Delta N^1\pm i \Delta N^2\,,~\Delta\bar{h}_{31}\mp i \Delta\bar{h}_{32}\bigr\}$:
\begin{align}
 &\omega^3
  +i\Bigl[ \frac{1}{\tau}
     +\frac{1}{\rho_0}\Bigl(\zeta+\frac{4}{3}\gamma\Bigr)k^2\Bigr] \omega^2
  -\frac{1}{\rho_0}\Bigl[ \frac{1}{\tau}\Bigl( \zeta+\frac{4}{3}\gamma\Bigr)
     +\Bigl(\frac{1}{\kappa}+\frac{4}{3}\mu\Bigr)\Bigr] k^2\omega
  -\frac{i}{\rho_0 \kappa\tau}\,k^2 = 0   \\
 &\mbox{~~~~~~\underline{fluid limit ($\tau\to 0$)}\,:~~}
   \omega^2
    +\frac{i}{\rho_0}\Bigl( \zeta+\frac{4}{3}\gamma\Bigr) k^2\omega
    -\frac{1}{\rho_0 \kappa}\,k^2 = 0 \\
 &\mbox{~~~~~~\underline{elastic limit ($\tau\to \infty$)}\,:~~}
   \omega^2
    +\frac{i}{\rho_0}\Bigl( \zeta+\frac{4}{3}\gamma\Bigr) k^2\omega
    -\frac{1}{\rho_0}\Bigl(\frac{1}{\kappa}+\frac{4}{3}\mu\Bigr)\,k^2 = 0 
\end{align}

\noindent
\textbullet~\underline{helicity 1 (sound modes)} 
$\bigl\{\Delta N^3\,,~\Delta\bar{h}_{33}\,,~\Delta\bar{h}_{11}+\Delta\bar{h}_{22}\bigr\}$:
\begin{align}
 &\omega^2
  +i\Bigl( \frac{1}{\tau} + \frac{\gamma}{\rho_0} k^2 \Bigr) \omega
  -\frac{1}{\rho_0} \frac{\gamma+\mu\tau}{\tau} k^2 = 0 
 \label{telegraph}\\
 &\mbox{~~~~~~\underline{fluid limit ($\tau\to 0$)}\,:~~}
   i\omega
    -\frac{\eta}{\rho_0} k^2 = 0 \\
 &\mbox{~~~~~~\underline{elastic limit ($\tau\to \infty$)}\,:~~}
   \omega^2
    +i\frac{\gamma}{\rho_0} k^2\omega
    -\frac{\mu}{\rho_0} k^2 = 0 
\end{align}

\noindent
\textbullet~\underline{helicity 2 (scalar modes)} 
$\bigl\{\Delta\bar{h}_{11}-\Delta\bar{h}_{22} \mp 2i\,\Delta\bar{h}_{12}\bigr\}$:
\begin{align}
 \omega=-\frac{i}{\tau}
\end{align}

\section{Conclusion and discussion}
In this work we have shown that viscoelasticity can have a universal 
description by introducing an intrinsic metric to express bonding structures 
and by adopting FPDs as an underlying gauge symmetry. 
We have seen that FPDs naturally interpolate between elasticity and fluidity, 
and that covariance under FPDs uniquely determines 
the dynamics in the first order of strains. 

The next step should be to extend our framework to the systems 
with more than one relaxation time and with multiple components, 
in order to describe more realistic materials.

We conclude with a comment that in the sound modes 
the diffusion equation becomes a telegraph-like equation 
when the relaxation time is taken finite (see Eq.\ (\ref{telegraph})). 
This means that information propagates only inside the ``light cone'', 
and thus the causality problem (inherent in diffusion equations) 
disappears when the effect of elasticity is taken into account.
This property should be taken over 
also in the ``relativistic extension of viscoelasticity'' \cite{fs}, 
and would serve as a causal completion of relativistic fluid mechanics.

\section*{Acknowledgments}
MF thanks the organizers of the QTS6 conference, 
especially Sumit Das and Al Shapere, 
for inviting him to give a talk.   
Elucidating discussions with Michael Berry and Yuho Sakatani are also appreciated. 
This work was supported by the Grant-in-Aid for the Global COE program 
``The Next Generation of Physics, Spun from Universality and
Emergence" from the Ministry of Education, Culture, Sports, 
Science and Technology (MEXT) of Japan. 
This work was also supported by the Japan Society for the Promotion of Science 
(JSPS) (No.\,20$\cdot$892) and by MEXT (No.\,19540288 and No.\,18540264).

\baselineskip=0.85\normalbaselineskip


\end{document}